\documentclass[12pt]{article}
\usepackage{graphicx}
\usepackage[scale=.79]{geometry}

\title{Fluctuations, Ghosts, and the Cosmological Constant}
\author{T. Hirayama and B. Holdom\\
\small\em Department of Physics, University of Toronto\vspace{-1ex}\\\small\em Toronto ON Canada M5S1A7}\date{}
\begin{document}
\maketitle
\begin{abstract}
For a large region of parameter space involving the cosmological constant and mass parameters, we discuss fluctuating spacetime solutions that are effectively Minkowskian on large time and distance scales.  Rapid, small amplitude oscillations in the scale factor have a frequency determined by the size of a negative cosmological constant. A field with modes of negative energy is required. If it is gravity that induces a coupling between the ghost-like and normal fields, we find that this results in stochastic rather than unstable behavior. The negative energy modes may also permit the existence of Lorentz invariant fluctuating solutions of finite energy density. Finally we consider higher derivative gravity theories and find oscillating metric solutions in these theories without the addition of other fields.
\end{abstract}

\section{Introduction}

One ingredient of the cosmological constant problem is the observation that the classical field equations of General Relativity do not admit a Lorentz invariant solution in the presence of a cosmological constant $\Lambda$. In this paper we will reconsider this point after relaxing the positive energy conditions on the energy momentum tensor. In particular we will consider the effect of negative energy contributions in the presence of a large $\Lambda$. We will suggest that it then becomes possible that the theory has solutions which are fluctuating on scales characterized by $\Lambda$ in such a way that no preferred frame is chosen, and which appear to be Minkowskian on larger length and time scales. That is, the dynamics controlled by a nonzero $\Lambda$ involves rapid fluctuations that are effectively averaged over by a low energy observer, and are such that no Lorentz symmetry breaking is induced in the low energy theory. For the moment we will simply comment on some consequences of such solutions, assuming they exist, keeping in mind that we will also be faced with issues associated with negative energy modes.

Although we are supposing that Lorentz invariance is preserved on high energy scales characterized by $\Lambda$, it is not explicitly preserved by nature on lower scales. It is broken by the energy momentum tensor $T_{\mu\nu}$ of whatever system we wish to study, and in particular it is broken by the $T_{\mu\nu}$ that defines the preferred cosmological frame. We would like to describe these explicit violations of global Lorentz symmetry as long wavelength perturbations to the high energy Lorentz invariant solution. If we consider performing a derivative expansion of the original coordinate invariant theory around the Lorentz invariant symmetry point, then this low energy theory up to two derivatives must be General Relativity. In order for it to be consistent with its interpretation as an expansion about a solution that is flat on large distances, the low energy theory must have vanishing cosmological constant.

We stress that given our assumptions, the vacuum energy component of $T_{\mu\nu}$ does not introduce explicit Lorentz symmetry breaking, whereas the remainder of $T_{\mu\nu}$ does. The cosmological constant influences the dynamics of the Lorentz invariant high energy solution, whereas the explicit Lorentz violating effects are accounted for in the low energy theory. Nevertheless, it seems paradoxical that the cosmological constant, which could be considered an infinite wavelength component of $T_{\mu\nu}$, has been absorbed by high energy physics. But this is a direct result of our assumption that $\Lambda$ influences the high energy theory by determining the amplitude of the Lorentz invariant fluctuations, and thus yielding solutions that are Minkowskian in the infrared. This is the flip side of the statement that the low energy theory does not contain a parameter $\Lambda$ which naively describes infinite wavelengths.

In section 2 we will illustrate the essential point with a model containing two decoupled fluids, with one of them having negative energy. The gravitational effects of these fluids will lead to a metric that is asymptotically flat in the presence of nonvanishing $\Lambda$. $\sqrt{|\Lambda|}$ still introduces a length scale in these solutions, just as it does for dS or AdS solutions, but here it sets the frequency of oscillation. The scale factor in the metric oscillates, as do the curvature invariants. The amplitude of this oscillation can vanish for special initial conditions, but typically it does not. The point about oscillating solutions is that their Minkowskian appearance on larger scales is a universal property of this whole space of solutions. Thus these solutions have an ultraviolet property, their frequency, and an infrared property, their effective flatness. The linking of these properties is at the origin of the counter-intuitive picture we are developing.

What is missing in these prototype solutions is Lorentz invariance, since the solutions are characterized by a preferred frame. To have a state that is both rapidly fluctuating and Lorentz invariant we need the Lorentz invariant analog of a rotationally invariant thermally fluctuating system. But unlike a rotation, a boost can generate states with arbitrarily large energies, and thus it appears that a nontrivial finite energy Lorentz invariant state cannot exist. We will argue that the situation can change in the presence of the negative energy modes, since they can serve to regulate the apparent infinite energies.

Of course negative energy modes, ghosts, raise the more immediate issue of stability, and so in section 3 we study interacting ghost dynamics using massive scalar field theory where the masses are large compared to the $\Lambda$ scale. A ghost instability can occur if the ghost couples to other fields, and so we explore the effect that gravity has when gravity is responsible for the coupling between the positive and negative energy modes. After including the effect of a changing scale factor on this coupling we find that solutions exhibiting unstable behavior are typically not present. In section 4 we consider a superposition of solutions involving different momentum modes and thus recover, at least formally, Lorentz invariance in the infinite mode limit. In section 5 we discuss a qualitative difference that occurs when a number of different fields are introduced, namely that the energy per mode need not be vanishing. As a classical model for Planck scale physics we study pure gravity theory involving higher derivatives in Section 6, where we find the analogs of the solutions in section 3. We conclude in section 7.

We emphasize that in this work we are using classical analysis to model physics that is likely close to the Planck scale. We do this because the dynamics of interest may well manifest itself in classical equations. This analysis misses the quantum instability associated with a ghost \cite{Carroll}, for example due to virtual graviton exchange if nothing else \cite{Cline}. But if the ghost mass is of order the Planck mass then the virtual graviton is far offshell in the deep Planckian region, and there is no guarantee that the perturbative quantum description holds. We assume that it does not.  Although our discussion in section 5 will lead us quite close to making some connection to quantum theory, we will leave this to pursue elsewhere.

\section{A Ghost Fluid}

We start with the notion of a ghost-like perfect fluid, completely decoupled from everything else except gravity. Whatever the constituents of this fluid are, their coupling to gravity will induce interactions between the fluid and other matter. We will return to the effect of these couplings in the next section, but for now we simply consider the effect of a ghost fluid on the gravitational field. We will only keep one degree of freedom of gravity, that of the scale factor $A(t)$ in the spatially flat FRW metric, $ds^2=dt^2-A(t)d\mathbf{x}^2$.

It is not surprising that a ghost fluid with negative energy and pressure and a normal fluid with positive energy and pressure could in combination cancel a cosmological constant, for some fixed value of the scale factor. We will demonstrate here that this can be a stable situation when the cosmological constant is negative. Namely, for deviations away from this solution, the solutions exhibit an oscillatory behavior for $A(t)$. The essential ingredients are a negative cosmological constant $\Lambda<0$, a ghost fluid with equation of state parameter $w_g>0$ (that is $p_g=w_g \rho_g$ with both $p_g$ and $\rho_g$ negative), and a normal fluid with, as we shall see, $0<w_n<w_g$.

Before solving the equations we describe the origin of the oscillation. If $A(t)$ oscillates then it has a maximum and a minimum, and at these points the total energy vanishes since it is proportional to $H(t)^2$ ($H=\dot{A}/2A$ is the Hubble parameter). Also at these points $p(t)\propto -\dot{H}(t)$, and thus the total pressure must be negative at the minimum and positive at the maximum. Starting at the minimum scale factor we have the energy $\rho_g+\rho_n+\Lambda=0$ and the pressure $p_g+p_n-\Lambda<0$; and since $\Lambda<0$, $\rho_n$ dominates $\rho_g$ and $p_g$ dominates $p_n$. Given that $w_g>w_n$, as the universe expands $|\rho_g|$ will fall faster than $\rho_n$ resulting in positive total energy. $|p_g|$ will fall faster than $p_n$ so that the total pressure becomes less negative and eventually changes sign as $\Lambda$ becomes relatively more important. Both $|\rho_g|$ and $\rho_n$ decrease relative to $\Lambda$, and after the pressure has changed sign, the total energy returns to zero, at which point the maximum scale factor is reached.

The ``energy'' and ``pressure'' Einstein equations, making use of the conservation of the separate fluids, are
\begin{eqnarray}
{3\over4}\,{\dot{A}(t)^{2}\over A(t)^2}&=&\Lambda-  A \left( t \right)  ^{-3/2-3/2\,w_{{g}}}r_{{g}}+  A \left( t
 \right)^{-3/2-3/2\,w_{{n}}}r_{{n}},\\
{\ddot{A}(t)\over A(t)}-{1\over4}{\dot{A}(t)^{2}\over A(t)^2}&=&
\Lambda+A(t)  ^{-3/2-3/2\,w_{{g}}}w_{g} r_{{g}}-A(t)  ^{-3/2-3/2\,w_{{n}}}w_{n} r_{{n}}.
\end{eqnarray}
Here $r_n$ and $r_g$ are positive constants. We start with a trivial time independent solution where for convenience we set $A(t)=1$. Then 
\begin{eqnarray}
r_g^0&=&-\Lambda(1+w_n)/(w_g-w_n),\nonumber\\
r_n^0&=&-\Lambda(1+w_g)/(w_g-w_n).
\label{e4}\end{eqnarray}
Since the $r$'s are positive we see again that $w_g>w_n$ if $\Lambda<0$. We also see that if $w_g$ is close to $w_n$ then the individual energy densities in the two fluids can be large compared to $\Lambda$.

We consider a small deviation from these values of the form 
\begin{eqnarray}
{r_g\over r_g^0}&=&1+{3\over2}(1+w_g)\varepsilon\nonumber\\
{r_n\over r_n^0}&=&1+{3\over2}(1+w_n)\varepsilon,
\end{eqnarray} 
where $\varepsilon$ is a small number of either sign. The following oscillating scale factor with amplitude $|\varepsilon|$ is then a solution, to first order in $\varepsilon$.
\begin{equation}
A(t)=1+(1-\cos(\omega t))\varepsilon
\label{e2}\end{equation}
The frequency is
\begin{equation}
\omega=\sqrt{-{3\over2}(1+w_g)(1+w_n)\overline{\Lambda}},
\end{equation}
where $\overline{\Lambda}=\Lambda$ at this order. Here we see why $\Lambda$ must be negative; a solution of the form (\ref{e4}) exists for $\Lambda>0$ and $w_g<w_n$, but it would not be stable under small deviations.

At second order in $\varepsilon$, we find that $\overline{\Lambda}$ is shifted
\begin{equation}
\overline{\Lambda}=\Lambda\, \left( 1+{\frac {3}{32}}\, \left( 2\,w_{{n}}+1-w_{{g}} \right)  \left( 2\,w_{{g}}+1-w_{{n}} \right)\varepsilon^2\right),
\end{equation}
and
\begin{eqnarray}
{r_n\over{r}_n^0}&=&1+{3\over2}(1+w_n)\varepsilon+{9\over16}(1+w_n)(w_n+w_g)\varepsilon^2,\nonumber\\
{r_g\over{r}_g^0}&=&1+{3\over2}(1+w_g)\varepsilon+{9\over16}(1+w_g)(w_n+w_g)\varepsilon^2.
\end{eqnarray}
The oscillating $A(t)$ picks ups a higher harmonic.
\begin{equation}
A(t)=1+(1-\cos(\omega t))\varepsilon+{1\over8}(w_n+w_g)\left[\cos(\omega t)-\cos(2\omega t)\right]\varepsilon^2
\end{equation}
The solution can be extended to order $\varepsilon^3$ by adding appropriate terms to $r_n$ and $r_g$, so that $A(t)$ picks up new terms proportional to $\cos(\omega t)$, $\cos(2\omega t)$, and $\cos(3\omega t)$ respectively. The existence of these solutions to all orders in $\varepsilon$ can easily be checked numerically.

For a given value of a negative $\Lambda$ there is no fine tuning of $r_g$ or $r_n$ necessary to obtain some solution in this class. The solutions are characterized by two properties, the averaged value of the scale factor and the amplitude of its oscillation, and these will adjust depending on the values of $r_g$ and $r_n$. (It was only for convenience that we expanded the solutions around $A(t)=1$.)

It would seem that we can simply average over these rapid fluctuations, leaving something that looks like Minkowski space on scales large compared to $1/\omega$. But the problem is that the effective theory below the $\Lambda$ scale is not General Relativity. The scale factor is oscillating around a fixed value determined by the energy densities in the conserved fluids, and so the scale factor degree of freedom no longer exists in the low energy theory. In other words this gravitational degree of freedom has picked up a mass, given by the frequency of oscillation in (\ref{e2}).

This problem is closely related to the loss of Lorentz invariance in the low energy theory, a result of considering fluids with a preferred rest frame in which only spatially constant solutions are considered. It is also related to our assumption that the fluids are noninteracting and separately conserved. But in fact interactions between these fluids are induced, a topic to which we now turn. In section 4 we return to the issue of Lorentz invariance.

\section{Ghost Interactions}

The previous discussion assumed that the ghost and normal fluids are completely decoupled. If we supppose that these fluids are composed of massive fields then gravity will induce a coupling between the fields, even at the classical level. This is because massive fields involve rapid time oscillations, and these oscillations feed into the metric through the gravitational coupling, and this in turn induces interactions between the positive and negative energy modes.

Before we consider this, let us first consider the effects of a nongravitational coupling between the ghost and normal fields. This coupled system can of course suffer from a severe instability, where the amplitudes of both fields can grow without bound while conserving total energy.
The nature of the instability depends on the form of the coupling between these fields \cite{Carroll}.
For some types of couplings the instability only occurs if the amplitudes of the fields and/or the size of the coupling is sufficiently large, while for other types of couplings the instability always occurs.

The simplest example of the latter case is when a ghost scalar $\phi_{g}$ and a normal scalar $\phi_{n}$ have a coupling $\phi_{g}\phi_{n}$, but are otherwise free fields. We will refer to these two real fields as the ghost and scalar respectively.
\begin{equation}
{\cal L}={1\over2}\partial_\mu \phi_n\partial^\mu \phi_n-{1\over2}m_n^2\phi_n^2
-{1\over2}\partial_\mu \phi_g\partial^\mu \phi_g+{1\over2}m_g^2\phi_g^2
+\lambda\phi_n\phi_g
\end{equation}
Note that the mass term of the ghost is also of ``wrong sign'' so as to ensure oscillatory solutions. For solutions of the form $\phi_{n}=\sigma(t)\cos(\mathbf{p}_{n}\cdot\mathbf{x})$, $\phi_{g}=\eta(t)\cos(\mathbf{p}_{g}\cdot\mathbf{x})$,
we look at the instability that occurs for equal masses $m_n=m_g=m$ between momentum modes that are in resonance $\mathbf{p}_{n}=\pm\mathbf{p}_{g}$. In terms of the resonant frequency $\omega=\sqrt{p^2+m^2}$ the solution is
\begin{eqnarray}
\sigma(t)&=&c_{1}\cos(\omega_{+}t+\delta_{1})\exp(\omega_{-}t)
+c_{2}\cos(\omega_{+}t+\delta_{2})\exp(-\omega_{-}t)\nonumber\\
\eta(t)&=&c_{1}\cos(\omega_{+}t+\delta_{1}-\kappa\pi/2)\exp(\omega_{-}t)
+c_{2}\cos(\omega_{+}t+\delta_{2}+\kappa\pi/2)\exp(-\omega_{-}t)\nonumber\\
2\omega_{-}^{2}&=&\sqrt{\omega^{4}+\lambda^{2}}-\omega^{2}\nonumber\\
2\omega_{+}^{2}&=&\sqrt{\omega^{4}+\lambda^{2}}+\omega^{2}\nonumber\\
\kappa&=&\mathrm{sign}(\lambda)\label{e1}
\end{eqnarray}
Thus there are exponentially increasing and decreasing solutions, and they differ in the relative phase of the oscillating factors in $\sigma(t)$ and $\eta(t)$. This feature will have implications in the next section.

These solutions are very model dependent. If there are other terms in the theory that become important for large field amplitudes then the exponential growth may not survive. For example adding terms of the form of a kinetic energy squared,  for one or both of the fields with appropriate signs, can change the exponential growth into an oscillating modulation of the high frequency oscillations. The amplitude of the modulation factor is inversely related to the size of coefficients of the higher derivative terms, and its frequency is proportional to $\lambda$, at least when this is small.

We now turn to a ghost+scalar system where the only coupling between the two fields is provided through their coupling to gravity. The two fields can only interact if $A(t)$ is nonconstant. Assuming the same form for $\phi_n$ and $\phi_g$ as before, the spatial dependence drops out of the two scalar field equations. After performing a spatial average of the ghost and scalar contributions to $T_{\mu\nu}$, the two field equations along with the ``pressure'' Einstein equation gives three equations for the three quantities $A(t)$, $\sigma(t)$, $\eta(t)$. The ``energy'' Einstein equation remains as a constraint.

Since the energy densities of the ghost and scalar have to be at least of order $\Lambda$, and the frequencies of these fields are of order $m\approx m_n \approx m_g$, then the field amplitudes are at least of order $\sqrt{|\Lambda|}/m$. In the following we will consider this ratio to be small. Our prejudice is that $\Lambda$ is not extremely small in Planck units, in which case $m$ can typically be of order $m_\mathrm{Pl}$. But this need not be the case; both $\sqrt{|\Lambda|}$ and $m$ can be much smaller than $m_\mathrm{Pl}$ and our description will still apply.

Our main interest is in the unequal mass case where it is clear that gravity influences the resonance condition $p_g^2+m_g^2=p_n^2+m_n^2$ for the interacting modes. The physical 3-momenta of the two modes are dependent on the scale factor, so that in terms of the comoving wavevector of each mode we have $p_g^2\rightarrow k_{g}^{2}/A(t)$ and $p_n^2\rightarrow k_{n}^{2}/A(t)$. Thus a changing scale factor, which is responsible for producing a coupling between ghost and scalar in the first place, can cause two modes $(k_{g},k_{n})$ that were in resonance to fall out of resonance.

Numerical analysis of the two mode case indicates that a stable, oscillating system can result as long as $k_{g}>k_{n}$.  $A(t)$ experiences oscillations determined by $\Lambda$, similar to (\ref{e2}). We refer to these as $\Lambda$ oscillations to distinguish them from the more rapid oscillations of the scalar fields. The basic stability is due to fact that the resonance condition for the fields is satisfied at only one value of the scale factor in the $\Lambda$ oscillation cycle. This can cause a disturbance to the motion, so that it is no longer strictly periodic, but the basic stability remains. The condition $k_{g}>k_{n}$ is equivalent to the previous condition $w_g>w_n$ in the two fluid model. The difference here is that we are taking the rapid time dependence of the massive fields into account, and including the back reaction on the metric. The latter is evident in the pressure equation where the matter terms oscillate and are largely canceled by an oscillating $\dot{H}(t)$ term. The ghost and scalar frequencies are affected by the changing $A(t)$ as it undergoes $\Lambda$ oscillations, and this leads to a beat type pattern in the sum of pressures, which in turn feeds back into $A(t)$. The $\Lambda$ oscillations that are evident in the individual ghost and scalar energies can be quite nonsinusoidal.

\section{Multi-mode solutions}

First consider a pair of modes, one positive and one negative energy, and each with a particular 4-momentum. A Lorentz transform of such a solution will also be a solution. For an arbitrarily large boost the positive and negative energies can be made arbitrarily large but they will nearly cancel, with the remainder canceling the cosmological constant just as before. For large energies and momenta the masses are less significant, and the ratios of energy to pressure of the positive and negative modes approach a common value. This is the analog of the situation described in (\ref{e4}) when $w_n\approx w_g$ and where the individual energy densities were also becoming large compared to $\Lambda$. 

For these Lorentz boosted solutions the amplitudes of the individual fields remain of order $\sqrt{|\Lambda|}/m$. If this is small then the solutions are close to being solutions to the linearized equations. Thus we can expect that a superposition of the Lorentz transformed solutions would also be a solution, up to small corrections. We can also add pairs of modes that are in resonance at differing values of the scale factor. The more pairs we add, the smaller the contribution of each to the total that cancels $\Lambda$.

\begin{figure}
\begin{center}\includegraphics[scale=.7]{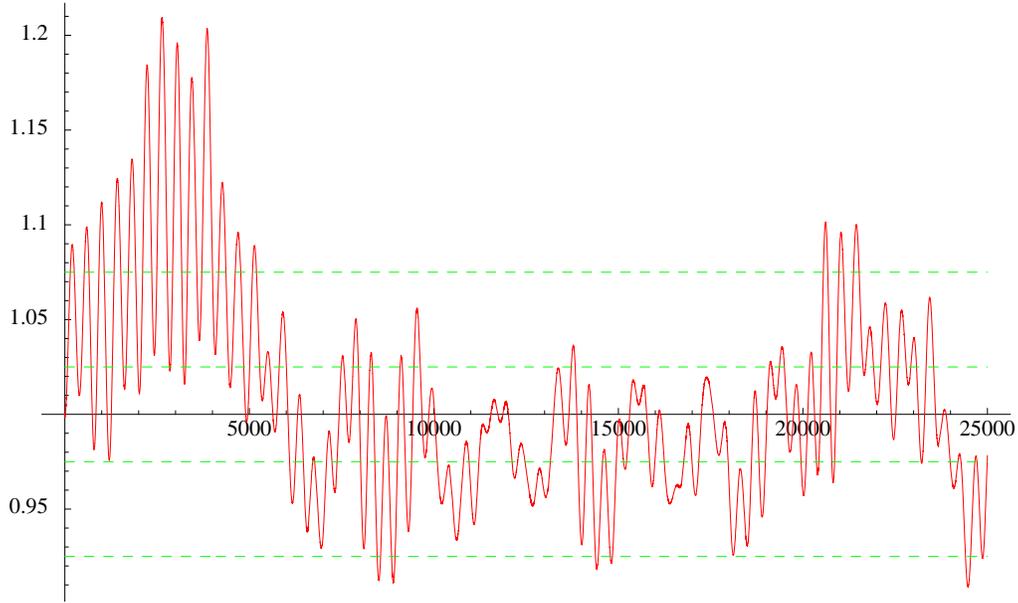}\end{center}
\vspace{-2ex}\caption{\small An example of a scale factor $A(t)$ for two pairs of positive and negative energy modes. The dashed lines show the values of $A(t)$ for which a resonance occurs. $m_n=1$, $m_g=.8$ and $\Lambda=.0001$ in Planck units. $A(t)$ also has a small amplitude, high (order 1) frequency component, but it is not visible.}
\end{figure}

If a large number of modes are excited then new pairs of modes can continually come into resonance as $A(t)$ changes. But as evident in (\ref{e1}), whether a pair of modes in resonance tend to increase or decrease in amplitude depends on their relative phases. If the phases of the various modes are effectively random, then the small changes in amplitudes caused by the short-lived resonances will also be random. We thus expect that the individual mode amplitudes display a stochastic behavior,\footnote{The system is of course deterministic for a given set of initial conditions.} and by the feedback on $A(t)$ the scale factor should also have a stochastic component.

We consider the case $m_n>m_g$, since this ensures that the pairs of modes in resonance satisfy $k_g>k_n$. We have numerically studied this case with up to 4 pairs of modes with varying values of $k_g$ and $k_n$. We observe the stochastic behavior along with the basic stability of the multi-mode system. Fig. (1) shows an example with two pairs of modes. The stability even survives the addition of direct interaction terms between the ghost and scalar, as long as the coupling is not too large.

We can contemplate taking the limit as the number of modes goes to infinity. The phase space is infinite, and here we are contemplating that this whole phase space is populated. In other words there would be positive and negative energy seas corresponding to a smooth distribution of modes covering the respective mass shells. We have seen how this can be built up from pairs of modes with nearly cancelling energies. In the limit, the energy contributed from each pair goes to zero, since the total energy density is finite.  That is, the energy density per unit phase space of these paired modes should vanish in the limit.

The interest in the infinite mode limit lies in its possible Lorentz invariance. The point is to have a situation where none of the fluctuating components, scalar, ghost, or metric, displays a preferred frame or 4-momentum. It is this that could guarantee that the low energy theory, obtained after integrating out the rapid fluctuations, is described by General Relativity.

If a collection of fluctuations is to display no preferred frame under Lorentz transformations, then we already know what its energy spectrum must look like. It must be proportional to the zero point energy spectrum of a field in a relativistic quantum field theory, where the energy per mode is ${1\over2}\hbar\omega_{\bf k}$. This type of classical background with all modes of the relativistic phase space excited, and with each mode having a random phase, has been considered in \cite{Boyer}. There it was explicitly checked that when a suitable average over the random phases is performed, Lorentz invariant results are obtained. In our case the amplitudes of the positive and negative energy modes must be such that the energy per mode is proportional to $\omega_{\bf k}^n$ and $-\omega_{\bf k}^g$ respectively, where $\omega_{\bf k}^i=\sqrt{{\bf k}^2+m_i^2}$. It is also possible for the energies to deviate from these values in some random way.

The main point for us is that the sum of positive and negative energy modes can be finite. We are suggesting that this finite result is due to the same dynamics that we have been able to study numerically for a finite collection of modes. This is the dynamics that responds to a finite $\Lambda$ to produce effectively flat space. In this way the infinite mode system is self-regulating. In fact it is very close in spirit to Pauli-Villars regularization, except that the negative energy states here are not fictitious. 

For a classical field theory without negative energy modes, the only Lorentz invariant configurations of finite energy density have constant fields. Here the negative energy modes allow the existence of less trivial configurations, which are the analogs of rotationally invariant states of fluctuating thermal systems. The obvious question then is why this unorthodox distribution over phase space doesn't cascade down and thermalize. The formal answer is Lorentz invariance; any thermalization implies that a preferred frame has been selected and thus a breaking of Lorentz symmetry. The unbroken symmetry prevents this from happening. For the same reason there can be no transfer of energy to or from the fluctuating background (for a related discussion see \cite{Boyer}).

\section{Many Fields}

We can analyze further how the normalization of the energy spectra is dynamically determined. We first consider our previous case of just one ghost field and one nonghost field with different masses. We introduce a momentum cutoff $M$ and consider the approach to the $M\rightarrow\infty$ limit. The ``energy'' and ``pressure'' components of the energy momentum tensor are
\begin{eqnarray}
 \Lambda+\int^M {d^3k\over(2\pi)^3}\, 
  \left( a_n\omega_{\bf k}^n -a_g\omega_{\bf k}^g\right) &=& 0
  \\
 -\Lambda+\int^M {d^3k\over(2\pi)^3}\, 
  \left( a_n\frac{|{\bf k}|^2}{3\omega_{\bf k}^n} 
   -a_g\frac{|{\bf k}|^2}{3\omega_{\bf k}^g}\right) &=& 0 .
\end{eqnarray}
The $a_i$'s determine the energy per mode normalization; we have omitted ${\bf k}$ subscripts because we have seen that they are ${\bf k}$ independent (at least on average) due to Lorentz invariance. These equations constrain $a_n$ and $a_g$ in such a way that they are vanishing as $M\rightarrow\infty$.
\begin{eqnarray}
 a_n&=&\frac{16\pi^2\Lambda}{(m_g^2-m_n^2) M^2} +{4\pi^2\Lambda+c\over M^4} +{\cal O}(M^{-6})
  \\
 a_g&=&\frac{16\pi^2\Lambda}{(m_g^2-m_n^2) M^2} -{4\pi^2\Lambda-c\over M^4} +{\cal O}(M^{-6}).
\end{eqnarray}
Thus the energy per mode is vanishing in the infinite mode limit, and this result is consistent with our numerical findings for this case.

But this result changes when more fields are involved. For $N$ fields we have
\begin{eqnarray}
 \Lambda+\int^M {d^3k\over(2\pi)^3}\, 
  \left(\sum_{i=1}^N a_i\omega_{\bf k}^i \right) &=& 0,
  \\
 -\Lambda+\int^M {d^3k\over(2\pi)^3}\, 
  \left(\sum_{i=1}^N a_i\frac{|{\bf k}|^2}{3\omega_{\bf k}^i} \right) &=& 0,
\end{eqnarray}
We expand the coefficients as follows
\begin{equation}
a_i=a_i^{(0)}+{a_i^{(2)}\over M^2}+{a_i^{(4)}\over M^4}+...
\end{equation}
where $a_i^{(2)}$ and $a_i^{(4)}$ are linear functions of $\ln(M)$. The equations can have solutions when
\begin{equation}
\sum_{i=1}^N a^{(0)}_i=\sum_{i=1}^N a^{(0)}_i m_i^2=\sum_{i=1}^N a^{(2)}_i=0.
\end{equation}
These three constraints are sufficient to cancel the quartic and quadratic divergences.\footnote{The elimination of the log divergences and the fixing of the finite pieces determines combinations of $a_i^{(2)}$ and $a_i^{(4)}$.} Thus to have \textit{nonvanishing} $a_i$ as $M\rightarrow\infty$ there must be at least one ghost and one nonghost and $N\geq3$. For larger $N$ we note that if there are at least three massive fields (involving both ghosts and nonghosts) then the remaining massless fields can have $a^{(0)}_i$'s that are equal in absolute value.

This shows how the energy per mode can stay finite in the $M\rightarrow\infty$ limit. In fact we would expect that all fields of the theory should be involved, due to the interactions between fields.  But as soon as we consider light fields, we are forced out of our classical discussion, since light fields are quantum mechanical. This in turn means that \textit{light} ghost fields are forbidden since they would introduce a severe quantum mechanical instability (or there needs to be a very low energy Lorentz violating cutoff \cite{Cline, Holdom}). On the other hand the classical Dirac theory describes negative as well as positive energy modes. As far as the classical analysis of this paper is concerned, the filling of the negative energy modes of this classical Dirac theory would serve the same purpose as the filling of the negative energy modes of a classical ghost theory. The essential difference is that the Dirac theory and its negative energy sea may survive the transition to quantum mechanics without the introduction of instabilities.

All of this leads to some speculations concerning the foundations of quantum field theory, with $\hbar$ as arising from the energy per mode normalization of classical fluctuations, which in turn is determined by $\Lambda$. We are pursuing this by calculating correlation functions from random phase classical fluctuations on a $1+1$ dimensional lattice, and comparing to the correlation functions of quantum field theory. We will report on these surprisingly positive results elsewhere \cite{us4}.
\newpage
\section{Pure Gravity}

Thus far we have used fluid models and scalar field theories to model ghost dynamics. In this section we would like to show how similar dynamics can arise if ghosts have a purely gravitational origin. We again seek solutions involving classical fluctuations of the metric around Minkowski space, with an amplitude determined by $\Lambda$. Of course such solutions do not exist in pure General Relativity. The required new massive degrees of freedom can appear when GR is supplemented with higher derivative terms. Higher derivative theories are well known to have ghosts with Planck scale masses, which are seen when linearizing the theory \cite{Stelle}. For a different perspective on these ghosts, see \cite{Hawking}.

As before, for the solutions of interest the amplitudes of the fluctuations will have to be of order $\sqrt{|\Lambda|}$ in Planck units, so that the energy can be of order $\Lambda$. We are able to find such solutions when $\Lambda$ is small and negative. Thus the basic result for pure gravity with negative $\Lambda$ is that besides the anti-de Sitter solutions, there are also solutions that are effectively Minkowskian at large distances. 

Our action is the general gravitational action with $4$ as well as $2$ derivative terms, including a cosmological constant.
\begin{equation}
S=\int\!{d}^{4}x\sqrt {-g}\left( \Lambda + \kappa R+a{R}^{2}+b R_{{\mu\nu}}R^{\mu\nu}\right)
\end{equation}
We choose to study the various physical degrees of freedom with a metric of the form
\begin{equation}
g_{{\mu\,\nu}}= \left[ \begin {array}{cccc} 1&0&0&0\\\noalign{\medskip}0&-A \left( t \right) &-B \left( t,z
 \right) &-B \left( t,y \right) \\\noalign{\medskip}0&-B \left( t,z \right) &-A \left( t \right) &-B \left( t,x
 \right) \\\noalign{\medskip}0&-B \left( t,y \right) &-B \left( t,x \right) &-A \left( t \right) \end {array}
 \right] 
\end{equation}
$$
B(t,x)=A(t)C(t)\sin(kx),\;\;\mathrm{etc,}
$$
such that $g_{\mu\nu}=\eta_{\mu\nu}+\delta g_{\mu\nu}$, where $\delta g_{\mu\nu}$ models the small amplitude fluctuations. Thus $A(t)$ is of order one but it contains an oscillating component describing a massive scalar mode. $C(t)$ is a graviton amplitude, containing both massless and massive degrees of freedom. The massive graviton is the ghost. The amplitudes of all these modes are small when $\Lambda$ is small. The masses of the scalar and massive graviton modes, as determined by the terms linear in these oscillations, are $m_{s}^{2}=-1/b$ and $m_{g}^{2}=1/(6a+2b)$ respectively in Planck units. The allowed range of $a$ and $b$ is then such that the mass squares are positive, but we otherwise take the parameters to be order unity. The parameter $k$ fixes the magnitude of the massive graviton 3-momentum.

Although the metric breaks rotational symmetry, there are remaining discrete symmetries under which $x$, $y$, and $z$ are interchanged which are convenient for simplifying the field equations. We also simplify the equations by making use of a small amplitude expansion for $C(t)$. A wave-type equation, linear in $C(t)$, can be obtained from the lowest order terms in the off-diagonal space-space components of the field equations. In the other equations the leading $C(t)$ dependence begins at quadratic order, and we consider the equations that result after a spatial averaging of these terms. This amounts to averaging the effective $T_{\mu\nu}$ from the graviton modes, and it leaves us with the energy and pressure equations with $k$ dependence. 

We can use the pressure equation and the graviton equation to numerically solve for $A(t)$ and $C(t)$. The energy equation should be satisfied identically during the evolution; this offers a good consistency check on the equations and the numerical routines. We provide initial conditions and then let the system evolve; if the system doesn't diverge or hit a singularity, and if this is true with no fine tuning of the initial conditions, then the solution can be considered classically stable. This is what we find for a class of oscillating solutions having a wide range of initial conditions---from one solution the initial conditions can be changed in quite arbitrary ways to obtain other solutions. $\Lambda$ is in the range $-\varepsilon <\Lambda<0$ where $\varepsilon \approx 0.0005$.

The rapid oscillations in $A(t)$ and $C(t)$ correspond to the masses of the scalar and graviton modes given above. The amplitude of these rapid oscillations is proportional to $\sqrt{|\Lambda|}$. There is also the expected longer period oscillation with a frequency of order $\sqrt{|\Lambda|}$. The amplitude of this oscillation in $A(t)$ is strongly dependent on the initial conditions. These features and the resulting stability is similar to what we described in the scalar field model. But there the ghost and scalar didn't interact except through a minimal coupling to gravity, and so it is interesting that the interactions between the massive scalar and graviton modes that occur in higher derivative gravity don't disrupt the oscillating solutions. We also note that the existence of the massless graviton doesn't disrupt the solutions.

The parameter $k$ of the massive graviton appears in the equations as $k^{2}/A(t)$, the square of the physical 3-momentum. This is constrained not to vanish. Since we have chosen the 3-momentum of the massive scalar mode to vanish, this is equivalent to our previous condition $w_g>w_n$. We can demonstrate the role of the massive graviton by removing it altogether, replacing it by some perfect fluid with equation of state $p=w\rho$, and leaving the massive scalar of higher derivative gravity. Now the metric is of pure FRW form. We find from numerical analysis that $p<0$, $\rho<0$, and $w>0$ is required, in agreement with the picture we have already developed. 

As another model we replace the massive graviton by a ghost-like massive scalar field with a nonminimal coupling $\xi \phi^{2}R^{2}$. $\xi$ adjusts the coupling between the positive and negative energy modes. Oscillating solutions exist once again for nonvanishing $\xi$. If we push $\Lambda$ down to the lower end of its range around $-0.005$, the amplitudes become large and the interaction between the modes becomes stronger. Here we find solutions having exponential expansion; these are the pure gravity analogs of the solutions discussed in the appendix of \cite{Holdom}.

\section{Conclusion}

We began by considering the effects of a nearly decoupled negative energy fluid, which in the presence of a negative cosmological constant and a positive energy fluid leads to prototype solutions in which the metric is fluctuating and is Minkowskian on large scales. We refer to these fluctuations as $\Lambda$ oscillations. We also used massive scalar fields to model the negative and positive energy modes; here the fields also oscillate due to their masses.  There is no fine tuning of the amplitude of these massive oscillations since the amplitude of the $\Lambda$ oscillations will adjust appropriately. The back reaction of the mass oscillations on the metric induces coupling between negative and positive energy modes. But since the value of a changing scale factor influences which pairs of modes are coupled at any given time, in the end this does not induce instability for a large range of parameters and initial conditions. Instead, a stochastic behavior develops.

Though the spacetime metric is Minkowskian on large scales, these prototype solutions lack Lorentz invariance. By considering a superposition of Lorentz
transformed solutions, we can formally obtain a Lorentz invariant analog
of a rotationally invariant thermally fluctuating system. The noncompact Lorentz group implies that an infinite superposition is required, but the consideration of such a superposition is not new \cite{Boyer}. In any case we conjecture that the cancellation of the cosmological constant survives this superposition, as we found for the superposition of a few pairs of positive and negative energy modes.  All these oscillating solutions require negative energy modes, and now the negative energy modes can be seen to be playing a second role in regulating the energy density of the infinite superposition. Under the assumption that the infinite superposition solution exists and is Lorentz invariant, the low energy theory should be General Relativity with vanishing cosmological constant.

General Relativity is the theory of Lorentz violating perturbations to this Lorentz invariant state. This includes all effects of high temperatures in the early universe, phase transitions, fields slowly rolling down potentials, periods of inflation, etc. But there may be several acceptable zero temperature ground states of the theory, corresponding for example to various local minima of some potential. In principle a low energy theory with an asymptotically flat spacetime could be set up around any of these minima, as long as the perturbations have energies less than the barrier energy separating the chosen vacuum from another.

We have proceeded beyond a scalar field theory description of ghosts in our study of higher derivative gravity theories. The higher derivative terms introduce additional Planck scale degrees of freedom, one of which is a massive ghost graviton.  No other fields are introduced and we see again the emergence of the prototype oscillating solutions. We expect that these types of solutions are typical of classical gravity theories with any number of higher derivatives. This discussion involves extremely massive negative energy modes, but in the context of light fields we also commented on the negative energy modes of the classical Dirac theory.

In this paper we have tried to demonstrate some links between Lorentz invariant classical fluctuations, the existence of negative energy modes, and the cosmological constant problem. But what does any of this have to do with the otherwise successful description of nature by a quantum field theory? We will report elsewhere on some intriguing results that may have some bearing on this question.

\section*{Acknowledgements}
This research was supported in part by the Natural Sciences and Engineering Research Council of Canada.

 \end{document}